%%%%%%%%%%%%%%%%%%%%%%%%%%%%%%%%%%%%%%%%%%%%%%%%%%%%%%%%%%%%%%%%%%%%
%  TeX Definitions                                                 %
%%%%%%%%%%%%%%%%%%%%%%%%%%%%%%%%%%%%%%%%%%%%%%%%%%%%%%%%%%%%%%%%%%%%

\newif\iffigs\figstrue
% Uncomment the next line if you do not want the figures
%\figsfalse

%
% the following is to use blackboard bold fonts --
\let\useblackboard=\iftrue
%
% activate this if you don't have them.
%\let\useblackboard=\iffalse
%
% You might also need to remove this line.
\newfam\black

\input harvmac.tex

\iffigs
  \input epsf
\else
  \message{No figures will be included.  See TeX file for more
information.}
\fi

\def\Title#1#2{\rightline{#1}
\ifx\answ\bigans\nopagenumbers\pageno0\vskip1in%
\baselineskip 15pt plus 1pt minus 1pt
\else%\special{papersize=11in,8.5in}%
\def\listrefs{\footatend\vskip 1in\immediate\closeout\rfile\writestoppt
\baselineskip=14pt\centerline{{\bf References}}\bigskip{\frenchspacing%
\parindent=20pt\escapechar=` \input
refs.tmp\vfill\eject}\nonfrenchspacing}
\pageno1\vskip.8in\fi \centerline{\titlefont #2}\vskip .5in}

\ifx\answ\bigans\def\tcbreak#1{}\else\def\tcbreak#1{\cr&{#1}}\fi
\useblackboard
\message{If you do not have msbm (blackboard bold) fonts,}
\message{change the option at the top of the tex file.}
\font\blackboard=msbm10 %scaled \magstep1
\font\blackboards=msbm7
\font\blackboardss=msbm5
%\newfam\black
\textfont\black=\blackboard
\scriptfont\black=\blackboards
\scriptscriptfont\black=\blackboardss
\def\Bbb#1{{\fam\black\relax#1}}
\else
\def\Bbb#1{{\bf #1}}
\fi
% *************************************
%
\def\yboxit#1#2{\vbox{\hrule height #1 \hbox{\vrule width #1
\vbox{#2}\vrule width #1 }\hrule height #1 }}
\def\fillbox#1{\hbox to #1{\vbox to #1{\vfil}\hfil}}
\def\ybox{{\lower 1.3pt \yboxit{0.4pt}{\fillbox{8pt}}\hskip-0.2pt}}
\def\np#1#2#3{Nucl. Phys. {\bf B#1} (#2) #3}
\def\pl#1#2#3{Phys. Lett. {\bf #1B} (#2) #3}

\def\comments#1{}

\def\half{{1\over 2}}

\def\a{\alpha}

\def\II{\relax{I\kern-.07em I}}

\def\IZ{\relax\ifmmode\mathchoice
{\hbox{\cmss Z\kern-.4em Z}}{\hbox{\cmss Z\kern-.4em Z}}
{\lower.9pt\hbox{\cmsss Z\kern-.4em Z}}
{\lower1.2pt\hbox{\cmsss Z\kern-.4em Z}}\else{\cmss Z\kern-.4em
Z}\fi}
\def\IB{\relax{\rm I\kern-.18em B}}

\def\ID{\relax{\rm I\kern-.18em D}}
\def\IE{\relax{\rm I\kern-.18em E}}
\def\IF{\relax{\rm I\kern-.18em F}}
\def\IG{\relax\hbox{$\inbar\kern-.3em{\rm G}$}}
\def\IGa{\relax\hbox{${\rm I}\kern-.18em\Gamma$}}
\def\IH{\relax{\rm I\kern-.18em H}}
\def\II{\relax{\rm I\kern-.18em I}}
\def\IK{\relax{\rm I\kern-.18em K}}
\def\IP{\relax{\rm I\kern-.18em P}}
%\def\IX{\relax{\rm X\kern-.01em X}}
%this doesn't work

\useblackboard
\def\IZ{\relax\Bbb{Z}}
\fi

\font\cmss=cmss10 \font\cmsss=cmss10 at 7pt
\def\IR{\relax{\rm I\kern-.18em R}}

\def\BZ{\IZ}

%%%%%%%%%%%%%%%%%%%%%%%%%%%%%%%%%%%%%%%%%%%%%%%%%%%%%%%%%%%%%%%%%%%%%%%%%%%%
%                    Definitions from LaTeX                                %
%%%%%%%%%%%%%%%%%%%%%%%%%%%%%%%%%%%%%%%%%%%%%%%%%%%%%%%%%%%%%%%%%%%%%%%%%%%%

%%%
%%% All those have problems with Font \rm
%%%

\def\lim{{lim}}

%%%%%%%%%%%%%%%%%%%%%%%%%%%%%%%%%%%%%%%%%%%%%%%%%%%%%%%%%%%%%%%%%%%%%%%%%%%%
%                    My definitions                                        %
%%%%%%%%%%%%%%%%%%%%%%%%%%%%%%%%%%%%%%%%%%%%%%%%%%%%%%%%%%%%%%%%%%%%%%%%%%%%
\input epsf

\def\SUSY#1{{{\cal N}= {#1}}}                   % N=? SUSY
\def\lbr{{\lbrack}}                             % [
\def\rbr{{\rbrack}}                             % ]

\def\wdg{{\wedge}}                              % wedge product

                              % Wilson lines

\def\inv#1{{1\over{#1}}}                              % inverse
                           % O(x)

               % Real numbers
               % Complex numbers

%%% \def\MR#1{{{\bf R}^{#1}}}               % Real numbers
%%% \def\MC#1{{{\bf C}^{#1}}}               % Complex numbers
               % Real numbers
               % Complex numbers
               % Circle, sphere,...
               % disk, ball,...
\def\MT#1{{{\bf T}^{#1}}}               % Torus
              % CP
               % Ruled surface F_n

\def\pch#1{{{\cal U}_{#1}}}             % Patch
\def\bdl{{{\cal L}}}                    % line-bundle
\def\px#1{{\partial_{#1}}}              % derivative

                 % Left large bracket
                % Right large bracket
\def\SLZ#1{{SL({#1},\BZ)}}              % SL(*,Z)

\def\hdg{{{}^{*}}}                      % Hodge star
                         % sign

%%%%%%%%%%%%%%%%%%%%%%%%%%%%%%%%%%%%%%%%%%%%%%%%%%%%%%%%%%%%%%%%%%%%%%%%%%%%
%                    Greek                                                 %
%%%%%%%%%%%%%%%%%%%%%%%%%%%%%%%%%%%%%%%%%%%%%%%%%%%%%%%%%%%%%%%%%%%%%%%%%%%%
\def\u{{\mu}}
\def\v{{\nu}}
\def\b{{\beta}}
\def\g{{\gamma}}

%%%%%%%%%%%%%%%%%%%%%%%%%%%%%%%%%%%%%%%%%%%%%%%%%%%%%%%%%%%%%%%%%%%%%%%%%%%%
%                    TITLE PAGE                                            %
%%%%%%%%%%%%%%%%%%%%%%%%%%%%%%%%%%%%%%%%%%%%%%%%%%%%%%%%%%%%%%%%%%%%%%%%%%%%

%%%\draftmode

%%%% PUPT-1672

%
\Title{ \vbox{\baselineskip12pt\hbox{hep-th/9612077, PUPT-1672}}}
{\vbox{
\centerline{A Note On Zeroes Of Superpotentials In F-Theory}}}
\centerline{Ori J. Ganor}
\smallskip
\smallskip
\centerline{Department of Physics, Jadwin Hall}
\centerline{Princeton University}
\centerline{Princeton, NJ 08544, USA}
\centerline{\tt origa@puhep1.princeton.edu}
%%%
\bigskip
\bigskip
\noindent
We discuss the dependence of superpotential terms in 4D F-theory 
on moduli parameters.  Two cases are studied: the dependence on
world-filling 3-brane positions and the dependence on 2-form VEVs.
In the first case there is a zero when the 3-brane hits the divisor
responsible for the superpotential.  In the second case,
which has been extensively discussed by Witten in 3D M-theory,
there is a zero for special values of 2-form VEVs when the M-theory
divisor contains non-trivial 3-cycles.  We give an alternative
derivation of this fact for the special case of F-theory.
 
\Date{December, 1996}

%%%%%%%%%%%%%%%%%%%%%%%%%%%%%%%%%%%%%%%%%%%%%%%%%%%%%%%%%%%%%%%%%%%%
%  B I B L I O G R A P H Y                                         %
%%%%%%%%%%%%%%%%%%%%%%%%%%%%%%%%%%%%%%%%%%%%%%%%%%%%%%%%%%%%%%%%%%%%

\lref\witfiv{E. Witten,
  ``Five-Brane Effective Action In M-Theory,''
  {\tt hep-th/9610234}}

\lref\witqtz{E. Witten,
  ``On Flux Quantization In M-Theory And The Effective Action,''
  {\tt hep-th/9609122}}

\lref\witnon{E. Witten,
      ``Non-perturbative Super-potentials In String Theory,''
      \np{474}{96}{343}, {\tt hep-th/9604030}}

\lref\svw{S. Sethi, C. Vafa and E. Witten,
      ``Constraints On Low-Dimensional String Compactifications,''
      \np{480}{96}{213}, {\tt hep-th/9606122}}

\lref\str{A. Strominger,
     ``Open p-Branes,'' \pl{383}{96}{44}, {\tt hep-th/9512059}}

\lref\vf{C. Vafa,
  ``Evidence For F-Theory,'' \np{469}{996}{403}, {\tt hep-th/9602022}}

\lref\GriHar{P. Griffiths and J. Harris,
  {\it Principles of Algebraic Geometry}, Wiley-Interscience, New York,
  1978.}

\lref\lozano{Y. Lozano,
  ``S-Duality in Gauge Theories As A Canonical Transformation,''
  \pl{364}{95}{19}, {\tt hep-th/9508021}}

\lref\BJPSV{M. Bershadsky, A. Johansen, T. Pantev, V. Sadov and C. Vafa,
  ``F-theory, Geometric Engineering and  N=1 Dualities,''
  {\tt hep-th/9612052}}

%%%%%%%%%%%%%%%%%%%%%%%%%%%%%%%%%%%%%%%%%%%%%%%%%%%%%%%%%%%%%%%%%%%%
%  P A P E R                                                       %
%%%%%%%%%%%%%%%%%%%%%%%%%%%%%%%%%%%%%%%%%%%%%%%%%%%%%%%%%%%%%%%%%%%%

%==================================================================%
%  Introduction                                                    %
%==================================================================%
\newsec{Introduction}
Super-potential terms in compactifications of M-theory to 3D on
a Calabi-Yau 4-fold $X$
arise from 5-branes wrapping non-trivial 6-cycles $D\subset X$
which satisfy \witnon:
\eqn\chid{
\sum (-)^p h^p(D) = 1.
}
If $X$ is elliptically fibered over the 3-fold $B$
one can take the F-theory limit
of zero size fibers \vf\ and find a super-potential in the 4D, $\SUSY{1}$
theory from a 3-brane that wraps a 4-cycle $E$ inside the base $B$.
It was shown in \svw\ that  in 3D M-theory compactifications there are
\eqn\numfil{
n_{2-branes} = \inv{24}\chi(X)
}
world-filling 2-branes which sit at a point on $X$
(if $\chi$ is not divisible by $24$, it was shown in \witqtz\ that
one has to add 4-form field-strengths in M-theory which modify this number).

The super-potential $W$ is a sum of terms corresponding to
5-branes wrapping divisors in $X$ which satisfy \chid\ \witnon.
Each separate term is of the form
\eqn\fevd{
f(\cdots) e^{-V_D + i\phi_D}
}
where $V_D$ is the volume of the divisor and
\eqn\phid{
\phi_D = \int_D \widetilde{C}.
}
Here $\widetilde{C}$ is the 6-form dual of the 3-form of M-theory.
The remaining pre-factor $f(\cdots)$ depends holomorphically on
the moduli and should be found from a 1-loop computation -- integrating
all the massive fields on the world-volume of the 5-brane.

The moduli on which $f$ could depend are thus:

\item{a.} 
The K\"ahler class of $X$
which joins the 3D duals of $C$ integrated on 2-cycles of $X$ to form
a complex scalar,

\item{b.} The complex structure of $X$,

\item{c.} The $h^{2,1}+h^{1,2}$ scalars which come from $C$,

\item{d.} The positions of the world-filling 2-branes.

In the F-theory limit, one takes the fibers to zero size.
M-theory then ``grows'' an extra decompactified dimension, the KK
modes of which are the almost massless states of 2-branes wrapped on
the small fiber.
In F-theory the superpotential is produced
from 3-branes wrapping a (4D) divisor $E$ in $B$ whose pull-back
to a (6D) divisor in $X$ satisfies \chid\ \witnon. The superpotential
term is still given by an equation similar to \fevd:
\eqn\feve{
f(\cdots) e^{-V_E + i\phi_E}
}
 but $V_E$ is now the volume of $E$ and
\eqn\phif{
\phi_E = \int_E B_4,
}
where $B_4$ is the self-dual 4-form of type-IIB.
The world-filling 2-branes become world-filling 3-branes \svw.
In the list above,  positions of 2-branes become positions of 3-branes
on $B$.
The modes of $C$ become the modes of the
two RR and NSNS 2-forms $B_{\u\v}^{(RR)}, B_{\u\v}^{(NS)}$.

In this paper we will discuss the behavior of the super-potential as
a function of the positions of the 3-branes and of the modes
of the two-forms.

In the first case, we will find that $f$ has a simple zero when the
3-brane hits the divisor $E$. The second case is a special case
of \witfiv. In 3D M-theory compactifications, the prefactor $f$ in
\witfiv\ arises from the partition function of the 5-brane world-volume
theory and the dependence on $C$ arises from the ``$dB\wdg C$''
interaction on the 5-brane world-volume.
In \witfiv\ the method for calculating the partition function for generic 
$D$'s was derived. For generic $D$'s, the difficulty is that there
is no manifestly covariant Lagrangian for the chiral 2-form which
lives on the 5-brane.
In the F-theory limit, when $D$ is elliptically fibered with fibers
of zero size, the 5-brane world-volume theory reduces (roughly)
to $\SUSY{4}$ $U(1)$ Yang-Mills theory with a variable coupling constant
and it is possible to calculate the partition function directly.
We will do that in section (3) and verify that it is a section 
of the requisite bundle \witfiv.

%==================================================================%
%  Dependence on 2-brane positions                                 %
%==================================================================%
\newsec{Dependence on 2-brane positions}
We will start by analyzing the dependence on the 3-brane positions.
For simplicity and more generality we will discuss the situation 
for 2-branes in M-theory.

Let us fix the positions of all 2-branes but one. The moduli space
for that single 2-brane is a copy of $X$.
We also keep all other irrelevant moduli frozen.
At first sight, since $X$ is compact and $f$
has no obvious source of singularities (as we will soon see,
a 2-brane that hits $D$ doesn't produce a pole), we might think
that $f$ should be constant. That conclusion, however, would be wrong
since $f$ is a section of a {\it non-trivial} line-bundle $\bdl$
over the moduli space $X$. 
To see this we will examine $e^{i\phi_D}$. As noted in \witnon,
$\phi_D$ is defined only up to a constant. 
As we change the position of the 2-brane to plot a small loop
around $D$ (the real codimension of $D$ in $X$ is 2) $\phi_D$
increases by $2\pi$. To define $\phi_D$
in this patch of moduli space we deform $D$ to $D'$ which doesn't intersect
the loop, pick $M$ such that $\partial M = D-D'$ and set:
\eqn\setphi{
\phi_D = \int_M\hdg dC.}
The 2-brane is a source for $C$ and
the loop intersects $M$ once. Thus, after the 2-brane has finished the
loop $\phi_D$ will pick up the flux from the 2-brane.

Geometrically, this means that $e^{-V_D + i\phi_D}$ is a section of 
the line bundle $\lbr -D\rbr$ associated with the divisor $D$.
(This line bundle is defined as follows \GriHar:
We take a neighborhood of $D$ as one patch and the complement of $D$
as the other patch. Let $g=0$ be a local defining equation for $D$,
then $1/g$ is the transition function of the line-bundle when going
from the complement of $D$ to the neighborhood of $D$.)
It follows that $f$ is a section of $\bdl=\lbr D\rbr$.
A holomorphic section of $\lbr D\rbr$ which has no poles must have
a simple zero on an analytic manifold homotopic to $D$. If we assume
that $D$ is isolated ($h^{3}(D)=0$) for example,
we find that $f$ is zero everywhere on $D$. 

We have seen that geometrically it is natural to expect a zero
when a 2-brane hits a 5-brane, but where does this zero come from,
physically?

%------------------------------------------------------------------%
% What happens when a 5-brane hits a 2-brane transversely?         %
%------------------------------------------------------------------%
\subsec{What happens when a 5-brane hits a 2-brane?}

We will argue that the zero comes from extra fermionic variables
that live on the 5-brane. In general they are massive, but when
a 2-brane sits on the 5-brane they become 2 massless anti-commuting
variables that are {\it localized} at a point on the 5-brane.
When one integrates them out one recovers the zero.

To argue the existence of these variables we take a (Euclidean)
5-brane  spread in directions $5,6,7,8,9,10$ and a 2-brane
in directions $0,1,2$. Compactifying the 10th direction we get a 
4-brane and a 2-brane of type-IIA. Since they don't have any
common `time' we cannot use standard D-brane techniques, but we
can `grow' a time direction by compactifying and T-dualizing along
the 4th direction, say.

 We obtain a 5-brane and a 3-brane
 that intersect along the common direction 4.
We have to find the  {\it zero modes} along this common direction.
We can now perform a Wick rotation and call that common direction
`time'. From the DD-strings we find two massless fermionic fields
on the $0+1$ intersection.
The fermions are charged under the difference of the $U(1)$-s
that live on the two branes, but on the $0+1D$ intersection
all that remains is $A_0$. So this gives a term
$$
i A_0 \psi_1 \psi_2
$$
There is also a Yukawa coupling 
$$
\Phi \psi_1\psi_2
$$
where $\Phi$ is the separation in the 3rd direction.
After T-duality back to  the 4-brane and 2-brane, the Wilson loop $A_0$
becomes the separation in the 4th direction and 
$\Phi+i A_0$  combines to a complex parameter $z$ that measures
the displacement in directions $3,4$.
The zero-modes of $\psi(x_4)$ give two anti-commuting {\it variables}
(with no coordinate dependence) that live at a {\it point} on the
5-brane and enter the Lagrangian via:
$$
z \psi_1 \psi_2.
$$
Integration over $\psi_1$ and $\psi_2$ produces the pre-factor $z$
which is the zero of the super-potential term.

%------------------------------------------------------------------%
%  Other D-brane intersections                                     %
%------------------------------------------------------------------%
\subsec{Other examples of localized variables}

We will briefly discuss two more cases where such localized variables
can appear.
The first is a world-filling 3-brane that hits a string world-sheet
instanton.
In M-theory, an instanton that corrects the K\"ahler metric can come from
a membrane that wraps a 3-cycle. The corresponding instanton in F-theory
would be a $(p,q)$-string whose boundary is on the 7-branes. The $(p,q)$
type would change according to the $\SLZ{2}$ transformations of F-theory
but the string is required to become elementary (i.e. $(1,0)$-type)
on the 7-branes. The embedding map cannot be holomorphic
 (since it intersects the 7-branes over a real dimension one), but
the area (in local string units) is required to be minimal.
A presence of a world-filling 3-brane near the stringy instanton
will add, as before, an extra localized variable. It is easy to analyze
this by making an $\SLZ{2}$ transformation to make the string a D-string.
As before, by T-duality this becomes a 2-brane intersecting a 4-brane
at a point in space and everywhere in time. This intersection breaks
supersymmetry but from the DD strings one finds again extra 
fermionic variables.

The other case is a world-filling 3-brane that approaches a type-IIB
$(-1)$-brane. In this case we can T-dualize to a 3-brane parallel to
a 7-brane. We find both fermionic and bosonic localized variables
which cancel each other, so there is neither a zero nor a pole.

%==================================================================%
%  Dependence on $C$                                               %
%==================================================================%
\newsec{Dependence on 3-form moduli}
In this section we will re-derive the result of \witfiv\ in a 
different way for the special case of an elliptically fibered
divisor $D$ with zero size fibers -- the F-theory limit.

%------------------------------------------------------------------%
%  Dependence on $C$                                               %
%------------------------------------------------------------------%
\subsec{Review of Witten's results}
We will start by reviewing some of the points from \witfiv.
The space of VEVs of $C$ is a hyper-torus of dimension $2h^{2,1}$.
Let
\eqn\thrf{
C = \sum_{i=1}^{2h^{2,1}(X)} a_i L_i,}
where 
\eqn\omegb{
L_i,\qquad i=1\dots H^3(X)
}
is a basis of integral 3-forms on $X$.

Each $a_i$ is periodic with
period $2\pi$. The complex structure on $T^{h^{2,1}}$ is given by
the decomposition into $(2,1)\oplus (1,2)$ forms.

The $U(1)$-bundle on $T^{h^{2,1}}$
of which $e^{i\phi_D}$ is a section is determined from
the kinetic term for $\phi_D$ in the 2+1D effective Lagrangian.
One finds \witfiv:
\eqn\kinphi{
\half (d\phi_D +{1\over 4\pi} \sum_{ij} Q_{ij}(D) a_i da_j)^2,
}
where $d$ is (2+1)-dimensional and the ``charges'' are given by
\eqn\qij{
Q_{ij}(D) = \int_D L_i\wdg L_j.
}
This term arises from the $C\wdg dC\wdg dC$ interaction of M-theory.

From \kinphi\ we see that the kinetic term for $\phi_D$ has a connection
term. The connection on the moduli space $T^{h^{2,1}}$ is given by
\eqn\contn{
A_i = -{1\over 4\pi}\sum_j Q_{ij}(D) a_i.
}
Thus, $e^{i\phi_D}$ is a section of a $U(1)$-bundle $\bdl^{-1}$
with first Chern 
class:
\eqn\fchc{
c_1(\bdl^{-1}) = -{1\over 4\pi}Q_{ij}(D) da_i\wdg da_j.
}
Since $D$ is an effective divisor, $\bdl$ is non-negative.

The pre-factor $f$ in \fevd\ will have zeroes on the moduli space,
when there are two integral 3-forms on $X$ whose
pull-backs to $D\subset X$ have a non-trivial intersection $Q_{ij}(D)$.
The zero should occur for special values of the integral of $C$ on
those 3-cycles.

The world-volume theory of the 5-brane is a 6D (twisted) tensor multiplet.
The 3-form couples to $B^{(-)}$ via an interaction that looks locally
like  $\int C\wdg dB$ \str.

In \witfiv\ the partition function was calculated by combining 
the anti-self-dual $B_{\u\v}^{(-)}$ with a self-dual part so that it
would be possible to write a covariant Lagrangian for them together.
The partition function is then a sum of terms which are sections 
of different line bundles over the moduli space (all have the 
same Chern class \fchc\ but they differ as complex line bundles).
Separating from the partition function that piece
which behaves as a section of the requisite line-bundle gives the 
required result.

%------------------------------------------------------------------%
%  F-theory                                                        %
%------------------------------------------------------------------%
\subsec{F-theory limit}
Let $D$ be an elliptic fibration with base $B$.
In the limit of zero-size fibers we have F-theory.
The 5-brane becomes a 3-brane wrapped on $B$.
The reduction of $B^{(-)}$ gives the $U(1)$ gauge field that
lives on a 3-brane.
To construct the partition function we need to divide $B$
into patches $\cup_\a \pch{\a}$. The patches are connected by
$SL(2,\BZ)$ transformations along the intersections $\pch{\a\b}$.
We will take the patches so that the intersections will be at the
boundary of the patches, i.e. $\pch{\a\b}$ will be 3-dimensional.

Locally on $B$ we pick a basis $(\xi_1^{(\a)},\xi_2^{(\a)})$
of integral 1-forms on the fiber such that they transform in the
fundamental $\SLZ{2}$ representation. Their Hodge duals inside the fiber
are
$$
{}^*\xi_1 = {1\over\tau_2} (|\tau|^2 \xi_2 + \tau_1 \xi_1),\qquad
{}^*\xi_2 = -{1\over\tau_2} (\xi_1 + \tau_1 \xi_2),
$$
where $\tau = \tau_1 + i\tau_2$ is the local modular parameter of the fiber.

The 3-form $C$ becomes two real 2-forms, the NS-NS and the RR 2-forms of
type-IIB:
\eqn\kbb{
C|_\pch{\a} = B^{NS,(\a)} \wdg\xi_1^{(\a)} 
              + B^{RR,(\a)} \wdg\xi_2^{(\a)}
}
which combine into a single complex 2-form $K$:
\eqn\combk{
K^{(\a)} = \tau_1^{(\a)} B^{NS,(\a)} 
    - i \tau_2^{(\a)}\,{}^*B^{NS,(\a)} - B^{RR,(\a)}.
}
Over the patch $\pch{\a}$ the action is
\eqn\actna{\eqalign{
I_\a =& \int_{\pch{\a}} \big\{ -{1\over 4g^2} (F-B^{NS})\wdg\hdg (F-B^{NS}) + 
{i \theta\over 32\pi^2} (F-B^{NS})\wdg (F-B^{NS}) \cr
   &-{i\over 2\pi} B^{RR} \wdg (F-B^{NS}) + fermions \big\}\cr
=& {1\over 4\pi}\int_{\pch{\a}} \left\{ -\tau_2 F\wdg\hdg F + 
i \tau_1 F\wdg F +2 i K \wdg F + \Lambda(K,\bar{K}) + fermions \right\},\cr
\Lambda(K,\bar{K}) =&
{1\over 4\pi}\int_{\pch{\a}}\left\{-\tau_2 B^{NS}\wdg\hdg B^{NS}
+ i \tau_1 B^{NS}\wdg B^{NS} + 2 i B^{RR}\wdg B^{NS}\right\},\cr
}}
where $\tau = {\pi i\over g^2} + {\theta\over 8\pi}$ (this is the normalization
for $U(1)$ as opposed to $SU(2)$)
 is the complex parameter of the elliptic fiber.

When passing from one patch to the other the background variables
are related by an $SL(2,\BZ)$ transformation:
\eqn\uaub{\eqalign{
\tau &\longrightarrow {a\tau + b \over c\tau + d},\cr
B^{NS} &\longrightarrow a B^{NS} + b B^{RR},\cr
B^{RR} &\longrightarrow c B^{NS} + d B^{RR}.\cr
}}
The gauge fields $A^{(\a)}$ and $A^{(\b)}$ are related by
an electric-magnetic duality transformation which is generically not
an algebraic relation. 
One has to add a piece $I_{\a\b}\{A^{(\a)}, A^{(\b)}\}$ to the action.
To find $I_{\a\b}$ we can take $\pch{\a\b}$ to be a plane at constant
time $t=0$. Then, $e^{i I_{\a\b}}$ would be related to the operator
that realizes electric-magnetic duality by \lozano:
\eqn\empsi{
\Psi'\{A^{(\a)}\} = \int \lbr DA^{(\b)} \rbr 
    e^{i I_{\a\b}\{A^{(\a)}, A^{(\b)}\}} \Psi\{A^{(\b)}\}.
}
Here $\Psi$ and $\Psi'$ are wave-functions before and after
electric-magnetic duality.

%%% with $ad-bc =1$.
One finds for $c\ne 0$:
\eqn\genk{
I_{\a\b}\{A,A'\} = 
\int_\pch{\a\b} \left\{ {a\over 2c} A'\wdg dA' 
               -{d\over 2c} A\wdg dA + {1\over c} A\wdg dA' \right\}.
}
For $c=0$ we have the $T^b$ transformations:
\eqn\cz{
e^{i I_{\a\b}\{A,A'\}} =
e^{{i b \over 2}\int_\pch{\a\b} A\wdg dA} \delta\{A-A'\}.
}
%%% The $\delta$-function insures that $A$ and $A'$ are equivalent up
%%% to a gauge transformation.
Finally we have to take care of the 2D triple intersections $\pch{\a\b\g}$.
First we have to make sure that the gauge transformations between patches
add up to integer multiples of $2\pi$ on $\pch{\a\b\g}$ and
second we have to add the two fermionic degrees of freedom which live
on the intersection of a 7-brane with a 3-brane.
To include the fermionic variables, we first choose the patches such
that all the singular fibers correspond to a monodromy $T\in SL(2,\BZ)$,
and then add:
\eqn\singul{
I_{sing} = \int d^2x\, \psi D_z \psi,
}
where $\psi$ is a chiral fermion that is charged with respect
to the gauge field $A$ of the patch.
The gauge anomaly of $I_{sing}$ is precisely what is needed to cancel
the anomaly from \cz\ for the 3D cut corresponding to the $T$
monodromy that has the 2D singular locus as a boundary.

Writing the partition function in full:
\eqn\parbc{
Z = Z_0 \int \prod_\a\lbr {\cal D}A^{(\a)} \rbr 
\lbr {\cal D}\psi \rbr
e^{
-\sum_\a I_\a\{A^{(\a)},K^{(\a)}\} 
   + i\sum_{\a\b} I_{\a\b}\{A^{(\a)},A^{(\b)}\}
    +  I_{sing}}.
}
If we switch back to Minkowskian metric we can find
the classical equations of motion for this action:
\eqn\clasol{
d({1\over g^2}F^{(\a)} +{\theta\over 8\pi^2}\hdg F^{(\a)}) = 0,
}
and on the 3D intersections $\pch{\a\b}$ we have
\eqn\match{\eqalign{
{1\over (g^{(\a)})^2}\vec{E}^{(\a)} 
+{\theta^{(\a)}\over 8\pi^2}\vec{B}^{(\a)} &= 
a\left({1\over (g^{(\b)})^2}\vec{E}^{(\b)} 
+{\theta^{(\b)}\over 8\pi^2}\vec{B}^{(\b)} \right) 
 + b \vec{B}^{(\b)},\cr
\vec{B}^{(\a)} &= 
c \left({1\over (g^{(\b)})^2}\vec{E}^{(\b)} 
+{\theta^{(\b)}\over 8\pi^2}\vec{B}^{(\b)} \right) 
+ d \vec{B}^{(\b)},\cr
}}
where $\vec{E},\vec{B}$ are the electric and magnetic field
on the 3D intersection.
We can combine the $F^{(\a)}$ data to an anti-self-dual 3-form $\omega$
on the 6D manifold $D$.
Writing locally
\eqn\lock{
\omega^{(\a)} = F^{(\a)}\wdg \xi_1^{(\a)} + 
(\tau_1 F^{(\a)} -i \tau_2{}^* F^{(\a)})\wdg\xi_2^{(\a)},
}
the $\omega^{(\a)}$'s join to form a closed anti-self-dual 3-form on $D$.
For a generic metric, the solutions to equations \clasol\ and \match\ 
will not respect the integrality conditions on $\pch{\a\b\g}$ and the
anti-self-dual 3-form $\omega$ will not come out integral.

%------------------------------------------------------------------%
% Transformation properties in $\bdl$                              %
%------------------------------------------------------------------%
\subsec{Transformation properties in $\bdl$}

To prove that $Z$ in \parbc\ is a section of the line-bundle $\bdl$,
we examine what happens as we change
\eqn\chgk{
C\longrightarrow C + 2\pi L
}
where $L$ is an integral 3-form on $D$.

We start with the simpler case of $D=B\times\MT{2}$.
There is only one patch and the integral reads:
\eqn\onepch{
Z = e^{\Lambda(K,\bar{K})}Z',\qquad
Z' = \int\lbr {\cal D}A \rbr e^{-\int\{\inv{4g^2}F\wdg\hdg F 
    + {i\theta\over 32\pi^2} F\wdg F + {i\over 2\pi} K\wdg F\}}.
}
Take $\chi$ to be an integral 2-form in $H^2(B,\BZ)$.
$Z'$ is invariant under 
\eqn\kkw{
K\longrightarrow K + 2\pi\chi.
}
The other period:
\eqn\kkgtw{
K\longrightarrow K + {2\pi^2 i\over g^2}(\hdg\chi) 
       + {\theta \over 8}\chi
}
does not leave $Z'$ invariant.
In fact, \kkgtw\ can be absorbed in a redefinition
\eqn\ffw{
F\longrightarrow F + 2\pi\chi,
}
which is allowed since $\chi$ is integral.
We see that up to terms that are independent of $K$ (denoted $C_0$)
we have
\eqn\zzw{
Z'(K + {2\pi^2 i\over g^2}(\hdg\chi)+ {\theta \over 8}\chi) 
   =  C_0 e^{i\int_B K\wdg\chi} Z'(K).
}
From \zzw\ it follows that $Z'$ is a section of the line bundle $\bdl$ 
and so is $Z$.

Now we take the more general case of variable fibers.
Assuming the patches are contractible, we expand
\eqn\lmxi{
L|_\pch{\a} = dM_1^{(\a)}\wdg \xi_1^{(\a)}
            + dM_2^{(\a)}\wdg \xi_2^{(\a)},
}
where $M_a^{(\a)}$ are local 1-forms on $B$.
Substituting \lmxi\ and \lock\ in $\int C\wdg dB$
we find that \chgk\ corresponds in the F-theory limit to
\eqn\chfl{
K^{(\a)}\longrightarrow K^{(\a)} + 2\pi \tau_1 dM_1^{(\a)}
        + 2\pi i\tau_2\, {}^*dM_1^{(\a)} + 2\pi dM_2^{(\a)}.
}
Note also that if we choose the patches so that all 7-branes 
are of $(1,0)$ type then $M_1$ vanishes on the 7-branes 
since the 1-cycle dual to $\xi_1$ vanishes on the 7-branes.

Now we change variables in \parbc:
\eqn\chgvar{
A^{(\a)}\longrightarrow A^{(\a)} + 2\pi M_1^{(\a)}.
}
The integrality of $L$ insures that the new $A^{(\a)}$ will satisfy
the integrality condition on $\pch{\a\b\g}$.
Combining this with \chfl, the action changes by (dropping $(\a)$):
$$
\Delta I_\a = \int \{ -i B^{NS}\wdg dM_2 + i F\wdg dM_2\}
$$
To get rid of $F\wdg dM_2$ we integrate it on $\pch{\a}$ to get
a contribution from the boundary only. 
This modifies $\pch{\a\b}$ but together with the change due to \chgvar\
and the relation between $M_{1,2}^{(\a)}$ and $M_{1,2}^{(\b)}$:
\eqn\mmsl{
M_1^{(\b)} = a M_1^{(\a)} + b M_2^{(\a)},\qquad
M_2^{(\b)} = c M_1^{(\a)} + d M_2^{(\a)}.
}
(since $M_1$ vanishes on the 7-branes there is no change in $I_{sing}$),
we find:
$$
Z \rightarrow
    e^{-i\sum_\a \int_\pch{\a} B^{NS,\a}\wdg dM_2^{(\a)}} Z.
$$
So, to sum up, if the 3-form $C$ on $D$ is given locally as
$$
C|_\pch{\a} = K_1^{(\a)} \wdg\xi_1^{(\a)} + K_2^{(\a)} \wdg\xi_2^{(\a)}
$$
and the integral 3-form $L$ on $D$ is given by
$$
L|_\pch{\a} = L_1^{(\a)} \wdg\xi_1^{(\a)} + L_2^{(\a)} \wdg\xi_2^{(\a)}
$$

The transition function of the bundle under $C\rightarrow C+2\pi L$ is
$$
e^{-i\sum_\a \int_\pch{\a} K_1^{(\a)}\wdg L_2^{(\a)}}
$$
Changing to the real coordinates \thrf\ on the moduli space:
$$
C = \sum_i a_i L_i,
$$
we see that under 
$$
a_i \longrightarrow a_i + 2\pi\delta_{ij}
$$
we have the transition function (a sum over patches is implicit):
\eqn\trfaj{
e^{i \sum_k a_k L_{2,k}\wdg L_{1,j}}.
}
We wish to show that \trfaj\ agrees with \fchc.
To this end we must find a gauge transformation 
$\Lambda(a_1\dots a_{2h^{2,1}})$ such that the connection
$$
A_i = -{1\over 4\pi}\sum Q_{ij} a_j + \px{i}\Lambda
$$
changes by \trfaj:
$$
A_k(a_1\dots a_j+2\pi \dots) - A_k (a_1\dots a_j\dots) =
   \sum_\a\int L_{2,k}\wdg L_{1,j}.
$$
Using
$$
Q_{jk} = \sum_\a \int (L_{2,k}\wdg L_{1,j} - L_{1,k}\wdg L_{2,j})
$$
We find
\eqn\lamsum{
\Lambda =
-\half\sum a_j a_k \int (L_{1,k}\wdg L_{2,j} + L_{2,k}\wdg L_{1,j}).
}

%==================================================================%
%  Discussion                                                      %
%==================================================================%
\newsec{Discussion}
Studying the dependence of superpotentials on the moduli
is important in order to determine whether supersymmetry is 
broken in a phase where a superpotential is generated.
We have seen that super-potential terms from 5-branes wrapping
divisors in M-theory develop a simple zero when a world-filling
2-brane sits on the divisor. The zero occurs because of 
the existence of extra light fermionic local variables on the 5-brane 
world-volume which appear when a 2-brane is close to the 5-brane.
When there are $n$ 2-branes sitting on the divisor
one finds a zero of order $n$. In order to restore super-symmetry,
one needs a zero of order two at least, so that the potential
which is proportional to the square of the derivative of the
superpotential will vanish.

In the second part of the paper we have verified that in the F-theory
limit, the dependence of the super-potential terms on 3-form VEVs
(which become RR and NS 2-forms in F-theory) agrees with the general
results of \witfiv. It would be interesting to try to generalize
the discussion of section (3) to calculate the partition function
of two coincident 5-branes on an elliptically fibered 3-fold.
This theory is related to the ``tensionless string theories'',
and \genk\ should somehow be generalized to a nonabelian S-duality
operator.

%==========================================================================%
% Acknowledgments                                                          %
%==========================================================================%
\bigbreak\bigskip\bigskip
\centerline{\bf Acknowledgments}\nobreak
I am very grateful to S. Ramgoolam and E. Witten for helpful discussions.
This research was supported by a Robert H. Dicke fellowship and
by DOE grant DE-FG02-91ER40671.

%==========================================================================%
% Note added                                                               %
%==========================================================================%
\bigbreak\bigskip\bigskip
\centerline{\bf Note added}\nobreak
As this work was completed, a paper \BJPSV\ appeared which
also discusses the extra localized variables
at the intersection of a Euclidean and a world-filling 3-brane.
In \BJPSV\ the variables were bosonic, but this is probably related
to the fact that the setting there was different and the
world-filling 3-brane was in a different phase.
The situation that was discussed in the present paper corresponds, in the
context of \BJPSV, to giving the quarks masses. In this phase
the superpotential has a zero when a quark mass vanishes.
The superpotential is then proportional to a fractional power of the quark
mass because the Euclidean 3-brane sits on $N+1$ 7-branes. In M-theory
this corresponds to a 2-brane on an $A_N$ singularity which
means that the transverse coordinate is the Nth root of the mass.

\listrefs
\end